\documentstyle[twoside,epsf]{article}

\catcode`\@=11
\long\def\@makefntext#1{
\protect\noindent \hbox to 3.2pt {\hskip-.9pt  
$^{{\eightrm\@thefnmark}}$\hfil}#1\hfill}		

\def\thefootnote{\fnsymbol{footnote}}
\def\@makefnmark{\hbox to 0pt{$^{\@thefnmark}$\hss}}	
	
\def\ps@myheadings{\let\@mkboth\@gobbletwo
\def\@oddhead{\hbox{}
\rightmark\hfil\eightrm\thepage}   
\def\@oddfoot{}\def\@evenhead{\eightrm\thepage\hfil
\leftmark\hbox{}}\def\@evenfoot{}
\def\sectionmark##1{}\def\subsectionmark##1{}}

\oddsidemargin=\evensidemargin
\renewcommand{\thefootnote}{\fnsymbol{footnote}}

\newcounter{sectionc}\newcounter{subsectionc}\newcounter{subsubsectionc}
\renewcommand{\section}[1] {\vspace{12pt}\addtocounter{sectionc}{1} 
\setcounter{subsectionc}{0}\setcounter{subsubsectionc}{0}\noindent 
	{\tenbf\thesectionc. #1}\par\vspace{5pt}}
\renewcommand{\subsection}[1] {\vspace{12pt}\addtocounter{subsectionc}{1} 
	\setcounter{subsubsectionc}{0}\noindent 
	{\bf\thesectionc.\thesubsectionc. {\kern1pt \bfit #1}}\par\vspace{5pt}}
\renewcommand{\subsubsection}[1] {\vspace{12pt}\addtocounter{subsubsectionc}{1}
	\noindent{\tenrm\thesectionc.\thesubsectionc.\thesubsubsectionc.
	{\kern1pt \tenit #1}}\par\vspace{5pt}}
\newcommand{\nonumsection}[1] {\vspace{12pt}\noindent{\tenbf #1}
	\par\vspace{5pt}}

\newcounter{appendixc}
\newcounter{subappendixc}[appendixc]
\newcounter{subsubappendixc}[subappendixc]
\renewcommand{\thesubappendixc}{\Alph{appendixc}.\arabic{subappendixc}}
\renewcommand{\thesubsubappendixc}
	{\Alph{appendixc}.\arabic{subappendixc}.\arabic{subsubappendixc}}

\renewcommand{\appendix}[1] {\vspace{12pt}
        \refstepcounter{appendixc}
        \setcounter{figure}{0}
        \setcounter{table}{0}
        \setcounter{lemma}{0}
        \setcounter{theorem}{0}
        \setcounter{corollary}{0}
        \setcounter{definition}{0}
        \setcounter{equation}{0}
        \renewcommand{\thefigure}{\Alph{appendixc}.\arabic{figure}}
        \renewcommand{\thetable}{\Alph{appendixc}.\arabic{table}}
        \renewcommand{\theappendixc}{\Alph{appendixc}}
        \renewcommand{\thelemma}{\Alph{appendixc}.\arabic{lemma}}
        \renewcommand{\thetheorem}{\Alph{appendixc}.\arabic{theorem}}
        \renewcommand{\thedefinition}{\Alph{appendixc}.\arabic{definition}}
        \renewcommand{\thecorollary}{\Alph{appendixc}.\arabic{corollary}}
        \renewcommand{\theequation}{\Alph{appendixc}.\arabic{equation}}
        \noindent{\tenbf Appendix \theappendixc #1}\par\vspace{5pt}}
\newcommand{\subappendix}[1] {\vspace{12pt}
        \refstepcounter{subappendixc}
        \noindent{\bf Appendix \thesubappendixc. {\kern1pt \bfit #1}}
	\par\vspace{5pt}}
\newcommand{\subsubappendix}[1] {\vspace{12pt}
        \refstepcounter{subsubappendixc}
        \noindent{\rm Appendix \thesubsubappendixc. {\kern1pt \tenit #1}}
	\par\vspace{5pt}}

\newcommand{\textlineskip}{\baselineskip=13pt}
\newcommand{\smalllineskip}{\baselineskip=10pt}

\def\eightcirc{
\begin{picture}(0,0)
\put(4.4,1.8){\circle{6.5}}
\end{picture}}
\def\eightcopyright{\eightcirc\kern2.7pt\hbox{\eightrm c}}

\def\abstracts#1#2#3{{
	\centering{\begin{minipage}{4.5in}\baselineskip=10pt\footnotesize
	\parindent=0pt #1\par 
	\parindent=15pt #2\par
	\parindent=15pt #3
	\end{minipage}}\par}} 

\renewenvironment{thebibliography}[1]			
	{\frenchspacing
	 \ninerm\baselineskip=11pt
	 \begin{list}{\arabic{enumi}.}
	{\usecounter{enumi}\setlength{\parsep}{0pt}
	 \setlength{\leftmargin 12.7pt}{\rightmargin 0pt} 
	 \setlength{\itemsep}{0pt} \settowidth
	{\labelwidth}{#1.}\sloppy}}{\end{list}}

\newcounter{itemlistc}
\newcounter{romanlistc}
\newcounter{alphlistc}
\newcounter{arabiclistc}

\newcommand{\fcaption}[1]{
        \refstepcounter{figure}
        \setbox\@tempboxa = \hbox{\footnotesize Fig.~\thefigure. #1}
        \ifdim \wd\@tempboxa > 5in
           {\begin{center}
        \parbox{5in}{\footnotesize\smalllineskip Fig.~\thefigure. #1}
            \end{center}}
        \else
             {\begin{center}
             {\footnotesize Fig.~\thefigure. #1}
              \end{center}}
        \fi}

\newcommand{\tcaption}[1]{
        \refstepcounter{table}
        \setbox\@tempboxa = \hbox{\footnotesize Table~\thetable. #1}
        \ifdim \wd\@tempboxa > 5in
           {\begin{center}
        \parbox{5in}{\footnotesize\smalllineskip Table~\thetable. #1}
            \end{center}}
        \else
             {\begin{center}
             {\footnotesize Table~\thetable. #1}
              \end{center}}
        \fi}

\def\@citex[#1]#2{\if@filesw\immediate\write\@auxout
	{\string\citation{#2}}\fi
\def\@citea{}\@cite{\@for\@citeb:=#2\do
	{\@citea\def\@citea{,}\@ifundefined
	{b@\@citeb}{{\bf ?}\@warning
	{Citation `\@citeb' on page \thepage \space undefined}}
	{\csname b@\@citeb\endcsname}}}{#1}}

\newif\if@cghi
\def\cite{\@cghitrue\@ifnextchar [{\@tempswatrue
	\@citex}{\@tempswafalse\@citex[]}}
\def\citelow{\@cghifalse\@ifnextchar [{\@tempswatrue
	\@citex}{\@tempswafalse\@citex[]}}
\def\@cite#1#2{{$\null^{#1}$\if@tempswa\typeout
	{IJCGA warning: optional citation argument 
	ignored: `#2'} \fi}}

\def\pmb#1{\setbox0=\hbox{#1}
	\kern-.025em\copy0\kern-\wd0
	\kern.05em\copy0\kern-\wd0
	\kern-.025em\raise.0433em\box0}

\def\fnt#1#2{\footnotetext{\kern-.3em
	{$^{\mbox{\scriptsize #1}}$}{#2}}}

\def\fpage#1{\begingroup
\voffset=.3in
\thispagestyle{empty}\begin{table}[b]\centerline{\footnotesize #1}
	\end{table}\endgroup}

\def\runninghead#1#2{\pagestyle{myheadings}
\markboth{{\protect\footnotesize\it{\quad #1}}\hfill}
{\hfill{\protect\footnotesize\it{#2\quad}}}}
\font\tenrm=cmr10
\font\tenit=cmti10 
\font\tenbf=cmbx10
\font\bfit=cmbxti10 at 10pt
\font\ninerm=cmr9

\font\eightrm=cmr8

\def\qed{\hbox{${\vcenter{\vbox{			
   \hrule height 0.4pt\hbox{\vrule width 0.4pt height 6pt
   \kern5pt\vrule width 0.4pt}\hrule height 0.4pt}}}$}}

\renewcommand{\thefootnote}{\fnsymbol{footnote}}	

\def\bsc{{\sc a\kern-6.4pt\sc a\kern-6.4pt\sc a}}	
\def\bflatex{\bf L\kern-.30em\raise.3ex\hbox{\bsc}\kern-.14em 
T\kern-.1667em\lower.7ex\hbox{E}\kern-.125em X} 
\def \be{\begin{equation}}
\def \ee{\end{equation}}
\def \bea{\begin{eqnarray}}
\def \eea{\end{eqnarray}}

\def \bx{{\bf x}}
\def \cH{{\cal H }}
\def \cA{{\cal A }}

\def \cS{{\cal S }}

\def \bk{{\bf k}}
\def \bq{{\bf q}}
\def \nd{{\vphantom{\dagger}}}

\def \yBCO6{{ YBa$_2$\-Cu$_3$\-O$_{7-\delta}$ }}
\def \yBCO6{{ YBa$_2$\-Cu$_3$\-O$_{6.6}$ }}
\def \yBCO6x{{YBa$_2$\-Cu$_3$\-O$_{6+x}$}}

\begin{document}

\runninghead{pSO(5)} {pSO(5)}

\normalsize\textlineskip
\thispagestyle{empty}
\setcounter{page}{1}


\vspace*{0.88truein}

\fpage{1}
\centerline{\bf Projected SO(5) Hamiltonian for Cuprates and Its Applications}
\vspace*{0.37truein}
\centerline{\footnotesize Assa Auerbach\footnote{assa@physics.technion.ac.il}  
and Ehud Altman\footnote{ehud@physics.technion.ac.il}}
\vspace*{0.015truein}
\baselineskip=10pt
\centerline{\footnotesize\it Physics Department, Technion }
\baselineskip=10pt
\centerline{\footnotesize\it Haifa 32000, Isreal, }
\baselineskip=10pt\centerline{\footnotesize\it and}
\baselineskip=10pt\centerline{\footnotesize\it Institute for Theoretical Physics,}
\baselineskip=10pt\centerline{\footnotesize\it University of California, 
Santa Barbara CA 93106}
\vspace*{0.225truein}

\vspace*{0.21truein}
\abstracts{The projected SO(5) (pSO(5)) Hamiltonian incorporates the quantum
spin and superconducting fluctuations of underdoped cuprates in terms of four bosons 
moving on
a coarse grained lattice. A simple mean field approximation can explain some key feautures
of the experimental phase diagram: (i) The Mott transition between antiferromagnet and 
superconductor, (ii) The  increase of $T_c$ and superfluid stiffness with
hole concentration $x$ and (iii) The  increase of antiferromagnetic  resonance
energy as $\sqrt{x-x_c}$ in the superconducting phase. We apply this theory to
explain the ``two gaps'' problem in underdoped cuprate SNS junctions. In particular
we explain the sharp subgap Andreev peaks of the differential resistance,
as signatures of
the antiferromagnetic resonance (the magnon mass gap).
A critical test of this theory is proposed.  The tunneling charge, 
as measured by shot noise, should change by increments of
$\Delta Q= 2e$ at the Andreev peaks, rather than by $\Delta Q=e$ 
as in conventional superconductors.
}{}{}



\pagebreak

\textheight=7.8truein
\setcounter{footnote}{0}
\renewcommand{\thefootnote}{\alph{footnote}}

\section{Introduction}
\noindent
The underdoped regime of high $T_c$ cuprate superconductors exhibits a pairing energy
(called pseudogap $\Delta_p$) which is much larger than the superconducting transition temperature  $T_c$.
$\Delta_p$ is
measured by NMR\cite{nmr}, tunneling\cite{tunn-exp}, photoemmission\cite{arpes} and other probes, as a sharp decrease
in  the single particle density of states below $\Delta_p$ at temperatures below $2\Delta_p$.
Assigning $\Delta_p$ to the local pairing energy, is consistent with the observed
anisotropy of $|\Delta_p(\bk)|$  in momentum space\cite{arpes} as expected for a 
$d$-wave symmetry
of the pair wavefunction.

In sharp contrast to conventional BCS superconductors,
$T_c$ and $\Delta_p$ are {\em not} proportional\cite{Deutscher}.
While $\Delta_p$ decreases with hole doping $x$, 
$T_c$ grows with $x$, as does the superfluid density $\rho_s$.
The latter relation is not coincidental as argued by Emery and Kivelson\cite{phase-fluc}.
Basically, since $\rho_s$ is small at low doping, superconductivity is
destroyed by
thermal phase fluctuations, and thus $T_c$ is determined by the Kosterlitz-Thouless
transition temperature $T_c \sim  \rho_s$, and not by pair breaking effects which are supressed 
at
$T_c<< 2\Delta_p$. 

The behavior of  $\rho_s \sim x$ is even more unconventional. It demonstrates that the correct number
of charge $2e$ bosons is proportional to the number of doped {\em hole pairs} away from the half filled 
Mott phase. The Mott phase exhibits long range antiferromagnetic (AFM) correlations.
This is a consequence  of  the strong local Hubbard interactions, which cannot
be easily handled as a weak coupling  perturbation of the electron gas.

The proximity of the AFM phase to the superconductor suggests that the effective Hamiltonian for the low
energy collective
modes would describe 4 bosons:  three magnons of the Heisenberg antiferromagnet, and 
the hole pairs. In fact, in addition there are gapless
fermionic particle-hole excitations near the nodes of $\Delta_P(\bk), \bk\approx (\pm\pi/2,\pm\pi/2)$.

In the AFM phase, one magnon branch condenses, yielding a finite staggered magnetization, while
the other two become the two spin wave modes. At higher doping, in the superconducting phase,  AFM order
disappears which allows these
modes to become massive with mass $\Delta_s$, and for optimally doped YBCO they  are observed by inelastic neutron scattering
as a resonance
at around 40meV near
the AFM wavevector. 

Such as model was introduced\cite{pSO5}  as the {\em ``projected SO(5) (pSO(5)) theory''}.
This model is constructed by projecting out doubly occupied configurations from
the local two-site SO(5) multiplets. The remaining  states are local singlets, 
triplets and hole pairs,
which define the
local bosons.

In the following, we review the pSO(5) model and its primary consequences by mean field  
theory. 
The important predictions are $\rho_s\sim x$, and that $\Delta_s(x) \sim \sqrt{x-x_c}$. 

We then use the pSO(5) model to predict current singularities in 
Superconductor-Normal-Superconductor (SNS)
junctions\cite{AA}, and propose
a future experimental test of this theory.
{\em The tunneling charges, as measured by shot noise,  should change by 
$2e$ at the Andreev peaks, rather than by $e$ as in conventional superconductors.}

\section{The Model}
\noindent 
The large onsite Hubbard repulsion between
electrons is  imposed by an apriori projection  of doubly occupied  states
from the Hilbert space\cite{comm-SO5}.

The undoped vacuum 
$|V\rangle$ is a half filled Mott insulator in a  quantum spin liquid state.
The pSO(5) vacuum possesses short range antiferromagnetic correlations.
 A translationally invariant realization of $|V\rangle$ on the microscopic
square lattice, is the short range resonating valence bonds (RVB)
state\cite{rvb,moshe}. Since local singlets without holes and double occupation can only
be produced with even number of sites,  we must choose a coarse grained lattice
in order to define the vacuum and the Fock space. 

If one defines a superlattice of dimers on the original square lattice, a Fock 
vacuum can be explicitly constructed in terms of electronic
operators as dimer singlets, where $i$ labels a dimer.
\be
|V\rangle_i={1\over \sqrt{2}} \left( 
c^\dagger_{i1,\uparrow} c^\dagger_{i2,\downarrow}
-c^\dagger_{i1,\downarrow}c^\dagger_{i2,\uparrow}\right) ~ |0)_i
\ee
where $|0)_i$ is an empty state of electrons.
The hole pairs  are simply
\be
b^\dagger_{h i }|V\rangle_i = ~|0)_i
\ee

The triplets are defined as bosonic excitations of the RVB vacuum i.e.
\bea
b^\dagger_{z i }|V\rangle_i &=& {1\over \sqrt{2}}\left( 
c^\dagger_{i1,\uparrow} c^\dagger_{i2,\downarrow}
+c^\dagger_{i1,\downarrow}c^\dagger_{i2,\uparrow}\right) ~|0)_i\nonumber\\
b^\dagger_{x i }|V\rangle_i &=& {1\over \sqrt{2}}\left( 
c^\dagger_{i1,\uparrow} c^\dagger_{i2,\uparrow}
+c^\dagger_{i1,\downarrow}c^\dagger_{i2,\downarrow}\right) ~|0)_i\nonumber\\
b^\dagger_{y i }|V\rangle_i &=&{1\over \sqrt{2}} \left( 
c^\dagger_{i1,\uparrow} c^\dagger_{i2,\downarrow}
-c^\dagger_{i1,\downarrow}c^\dagger_{i2,\uparrow}\right) ~|0)_i
\eea

Out of the undoped RVB
vacuum $|V\rangle$, $b^\dagger_{h}$ create charge $2e$ bosons (hole pairs) and $b^\dagger_{\alpha}$, $\alpha=x,y,z$
create a triplet of antiferromagnetic,  spin one  magnons.
For the square lattice, a  similar construction which preserves the lattice rotation group
and has explicit $d$-wave symmetry of the
hole pairs bosons, will be presented in a forthcoming publication\cite{ehud}.

The  lattice pSO(5) Hamiltonian is
\bea
\cH^{pSO(5)}&=&\cH^{charge} + \cH^{spin}+\cH^{int} +\cH^{Coul}+\cH^{ferm}
\nonumber\\
\cH^{charge} &=& (\epsilon_c- 2\mu) \sum_i b^\dagger_{h i } b^\nd_{ h i
} - {J_c\over 2} \sum_{\langle ij\rangle} \left( b^\dagger_{ h i } b^\nd_{ h i
}+\mbox{h.c.}\right)\nonumber\\ \cH^{spin} &=& \epsilon_s \sum_{i\alpha  }
b^\dagger_{ \alpha i} b^\nd_{ \alpha i} - {J_s} \sum_{\alpha \langle
ij\rangle} n^\alpha_i n^\alpha_j
\nonumber\\  \cH^{int} &=& W \sum_i
:\left( b^\dagger_{hi}b^\nd_{hi} +\sum_\alpha
b^\dagger_{ \alpha i}b^\nd_{ \alpha i} \right)^2 : , \nonumber\\
\label{pSO5}
 \eea
where $:():$ denotes  normal ordering, and
$n^\alpha_i=
(b^\dagger_{i\alpha}+b^\nd_{i\alpha})/\sqrt{2}$ is the  N\'eel spin field.
 $\cH^{int} $ describes  short range interactions between
bosons, and $\cH^{Coul}$
describes  the long range Coulomb interactions.  $H^{ferm}$ describes coupling
to the nodal (fermionic) quasiparticles, which  contribute a 
finite density of single electron states at low energies. We shall
 not compute the  fermion  contributions, and will discuss them in detail
 elsewhere\cite{ehud}.

\section{Mean Field Theory: Results}
\noindent
The uniform mean field approximation to
Eq. (\ref{pSO5}) is straightforward\cite{pSO5}. It amounts to replacing
$b_{\gamma  i} \to \langle b_{\gamma}\rangle$,
$\gamma=h,\alpha $. The order parameters are related to  
experimental observables: the Bose condensate of hole pairs is the 
superconducting order parameter

\be  
x=\langle b^\dagger_{h }\rangle = \langle \sum_{ij}d_{ij}c_{\uparrow i} c_{\downarrow j}\rangle  
\ee
where $d_{ij}$ is the normalized short range pair wavefunction with $d$ wave symmetry.
Bose condensation of magnons yields the staggered magnetization
\be  
y=\langle b_{z }\rangle = \langle
n^z\rangle 
\ee

The $T=0$ variational energy of
$\cH^{charge}+\cH^{spin}+\cH^{int}$  is 
\be 
E^{MFT}/{\cal N}= (\epsilon_c-2\mu-{z\over 2} J_c) x^2 +
(\epsilon_s-zJ_s) y^2 + W(x^2+y^2),
\ee
where $z=4$ is the square lattice coordination and ${\cal N}$ is the lattice size. 
Minimizing $E(x,y)$ we find a first order
transition between two phases:
\bea
\mu<\mu_c~~~&&\mbox{AFM insulator:} ~~x=0,y\ne 0\nonumber\\
\mu>\mu_c~~~&&\mbox{d-SC:} \nonumber~~x\ne 0, y=0
\eea
where 
\be
\mu_c = {1\over 2} \left( (\epsilon_c-\epsilon_s)- ({z\over 2}J_c -z J_s)\right)
\ee

At $\mu<\mu_c$ we have an  undoped Mott insulator  
with  no hole
pair bosons, and where the magnons Bose-condense. The condensate  supports  a
finite staggered magnetization 
\be
 |\langle
n^\alpha\rangle|^2 = (2J_s-{1\over 2}\epsilon_s)/W\equiv m_s^2   ~~~~~ \mu < \mu_c 
\ee
Expanding the mean field Hamiltonian to second order in the Bose operators,
one obtains two linear 
spin wave modes at
\be
\omega=c|\bq|, ~~~~~c= 2\sqrt{2} J_s/\hbar .
\ee
$c$ is
the semiclassical spinwave  
velocity which agrees with semiclassical limit of the Heisenberg antiferromagnet.

At  $\mu>\mu_c$ the ground state  becomes doped with hole pairs which
Bose-condense into a
superconducting phase with an order parameter
\be
|\langle b^\dagger_{h } \rangle|^2 = (\mu-\mu_c)/W +m^2_s~~~~~
\mu > \mu_c \label{Delta}
\ee

The mean field phase stiffness  is given by $\rho_c=J_c  \langle
b^\dagger_{i } \rangle^2$,   and therefore Eq. (\ref{Delta})  explains 
why $\rho_c$  increases with chemical potential (and doping) in the
underdoped superconducting regime, as observed  experimentally\cite{uemura}.
Quantum phase fluctuations which increase with $\rho_c^{-1/2}$ significantly
reduce the mean field order parameter (\ref{Delta}) near the transition.

Long range interactions in $\cH^{Coul}$, 
frustrate  the first order transition and create
intermediate (possibly incommensurate) phases\cite{pSO5}, which we shall not
discuss here.

Analysis of the linear quantum fluctuations about mean field theory 
in the superconducting phase, when $y=0$, yields three massive magnons (of spin 1).
Their mean field Greens function approximates the spin structure factor 
\be
S_{\alpha\alpha'}(\omega,\bq)\approx s_0
{\delta_{\alpha\alpha'} \over \omega^2-c^2(\bq-\vec{\pi})^2-\Delta_s^2 }
\label{Sqom}
\ee
Here $c$  is the spin wave velocity, and $s_0$ is a normalization factor.
The  poles of  Eq. (\ref{Sqom}) have a mass gap at $\Delta_s$.
The mean field magnon gap is found to depend on the chemical potential as follows:  
\bea
\Delta_s&=&2\sqrt{(\mu-\mu_c)(\mu-\mu_c  +2J_s)}\propto \sqrt{x-x_c}\nonumber\\
c(\mu)&=&2\sqrt{2} J_s\sqrt{1+(\mu-\mu_c)/(2J_s)} 
\eea
which by Eq. (\ref{Delta}) also implies that $\Delta_s^2$ increases,
and the magnon dispersion stiffens at higher doping.

Thus the pSO(5) mean field theory can explain the   systematic   increase
of $\Delta_s$ with $T_c$ which has been
observed  by Fong {\it et. al.}\cite{Fong}.
The doping dependent resonance energy $\Delta_s(\delta) $ increases\cite{Fong}
between $\Delta_s(0.5)= 25$meV, (with $ T_c=52^\circ K$),
and  $\Delta_s(1)= 40$meV, (at $T_c=92^\circ K$)\cite{Fong}.
\vspace{0.5cm}

\section{SNS Junction: KBT theory}
Peaks in the differetial resistance of Superconducting-Normal-Superconducting (SNS)
junctions
have been customarily interpreted using the  theory of
multiple Andreev reflections, following  Klapwijk, Blonder,
and Tinkham (KBT)\cite{kbt}.

\begin{figure}[htbp]
\vspace*{13pt}
 \leavevmode \epsfxsize0.75\columnwidth
\epsfbox{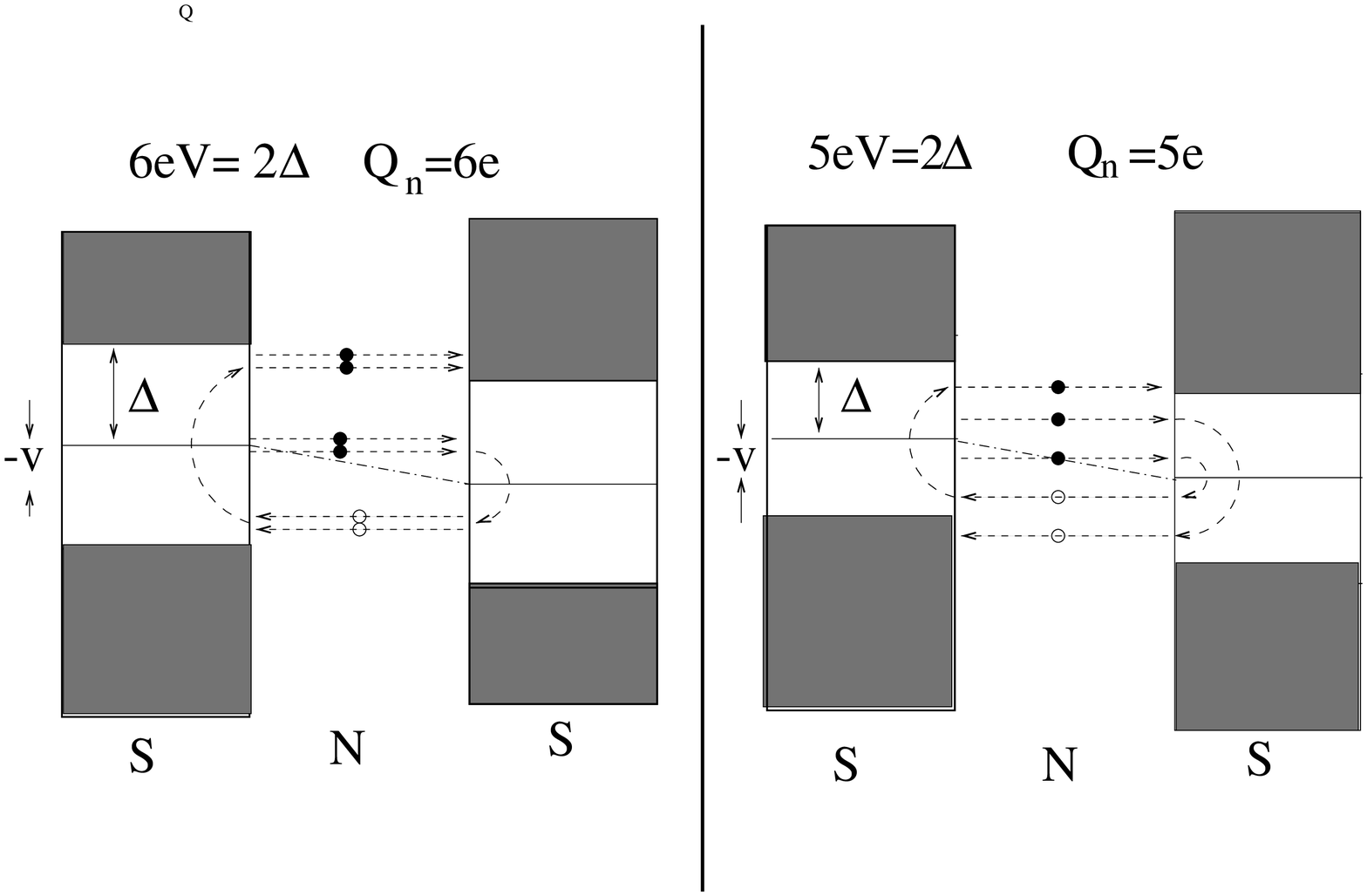}\vskip0.2pc
\fcaption{{\bf KBT Theory.} Differential resistance peaks of  $n=6$  (left
diagram), and $n=5$ (right diagram), involve  a cascade of $n$
Andreev reflected charges traversing the normal metal. Singular dissipation is
due to emission of  quasiparticles above the $s$-wave gap.
Filled (empty) circles denote electrons (holes) in the normal barrier.}\label{fig:btk}
\end{figure}

KBT theory treats two conventional superconductors
with a single $s$-wave BCS quasiparticle gap  $\Delta$,  separated by a free
electron metal. Electrons traversing the  metal are Andreev reflected back as
holes, gaining energy increments $eV$ at each traversal (as depicted in Fig.
\ref{fig:btk}).  Peaks in the differential resistance appear at voltages
$2\Delta/ne$, and are due to the $(E-\Delta)^{-1/2}$ singularity in the
quasiparticles'  density of states.
However, in cuprate SNS junctions, such as YBa$_2$\-Cu$_3$\-O$_{6.6}$  -
YBa$_2$Cu\-$_{2.55}$\-Fe$_{0.45}$\-O$_{y}$ -YBa$_2$\-Cu$_3$\-O$_{6.6}$
examined by Nesher and Koren\cite{nesher} , application of KBT theory is
problematic.   A naive  fit to KBT expression faces the two gaps puzzle, i.e.
an ``Andreev  gap'' is of order  $\Delta\approx 16$meV, while the
tunneling gap is about three times larger\cite{Racah}, and scales differently
with $T_c$.  Without perfect alignment of the interfaces, it is
hard to understand the observed sharpness of peaks\cite{nesher} since the
$d$-wave  gap is modulated at different directions. Moreover,  the barrier is
by no means a ``normal'' metal devoid of interactions:  it is
 an underdoped cuprate with  antiferromagnetic correlations and
strong pairing interactions as evidenced by a large proximity
effect\cite{prox}.

In the following sections we review an alternative explanation for
the differential resistance peaks series\cite{AA}, which takes into
account the strong correlations in the pseudogap regime. Our analysis  resolves
Duetscher's two energy scales puzzle\cite{Deutscher}.

\section{SNS Junctions: pSO(5) Theory}
\noindent
We consider a  junction, where  the barrier (N)  has no
superconducting  or magnetic order $\langle b^\dagger_h\rangle=0,\langle
n^\alpha\rangle=0$.  We  derive on general grounds  the form of the
effective tunneling Hamiltonian between superconductors as follows.

An integration of the barrier's charged bosons  $b_{h}$ out of the path
integral results in an effective action $\cS^{tun}$ which couples the charges
of the two  superconductors . $\cS^{tun}[b_{h_L},b_{h_R},b_\alpha]$
explicitly  depends on the hole pairs  bosons on  the left and right
interfaces, and on the magnons  in the barrier. By charge conservation,  an
expansion of $\cS^{tun}$  as a power series leaves only terms with equal
number of $b_h$'s and $b^\dagger_h$'s. By spin conservation, the magnon
terms are singlets and hence at least bilinear in 
$n^\alpha$.
\begin{figure}[htbp]
\vspace*{13pt}
    \leavevmode \epsfxsize0.75\columnwidth
    \epsfbox{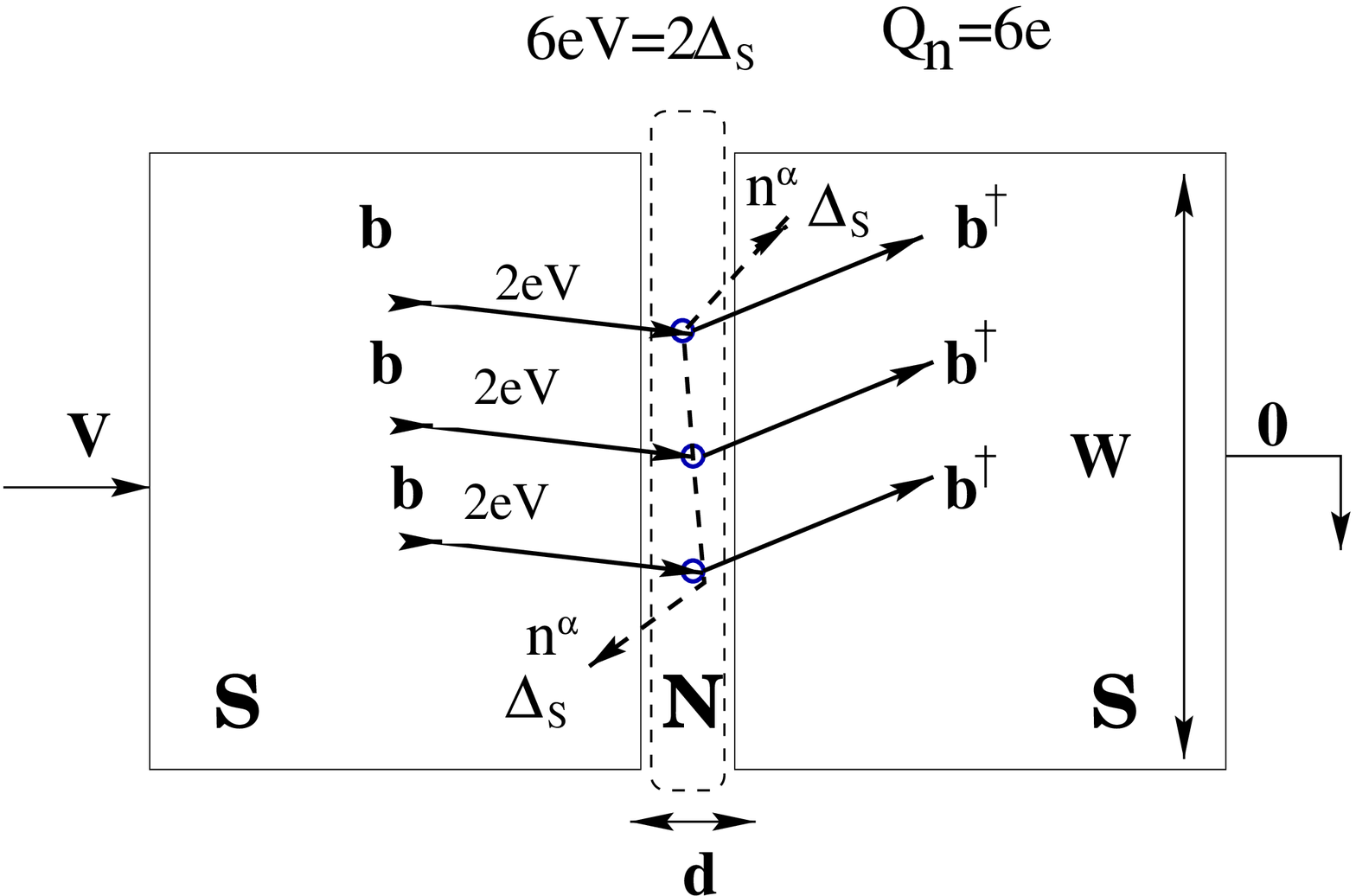}\vskip0.2pc

\fcaption{{\bf pSO(5) theory for Andreev peaks in cuprate SNS junctions}.
Three hole pairs co-tunneling from left to right, generate a pair of magnons.
At the antiferromagnetic resonance threshold $6eV=2\Delta_s$, this
process contributes to the $n=3$ peak of the differential resistance. The
diagram contains lowest order contributions of  hole pairs-magnon interactions
to the tunneling vertex $T_3$. } \label{fig:diag} 
\end{figure}

This expansion leads to a series of tunneling terms. For the Andreev peaks
we  retain  only the leading order terms (in $b^\dagger,b$) which are
\bea
 \cH^{tun-mag} &=& \sum {\delta^{2n+2} \cS\over \delta b_{h i_1} \ldots 
\delta b_{h i_{2n}} \delta n^\alpha \delta n^\alpha}~ 
 b_{h i_1} \ldots 
 b_{h i_{2n}}   n^\alpha   n^\alpha \nonumber\\
&=& -
\sum_{n} (\cA^\nd_n + \cA^\dagger_n) \nonumber\\ \cA_n&=& \sum_{y_1\ldots
y_{2n},\bx,\bx'} T_n  b^\dagger_{h_L, 1} \ldots b^\dagger_{h_L,n} ~b^\nd_{h_R
n+1} \ldots b^\nd_{h_R ,2n } \nonumber\\ && \times \left( \sum_\alpha
n^\alpha(\bx) n^\alpha (\bx') \right) \label{H-tun}
\eea
$\cA_n^\dagger$  describes  a simultaneous tunneling of $n$ hole
pairs from the left to the right superconductor,  coupled to
a magnon pair  excitation.  $T_n$ is the tunneling vertex function,
which depends on the bosons positions.  

 The energy transfer mechanism is depicted
diagrammatically in Fig.\ref{fig:diag}. We do not compute $T_n$'s
which depend on the details of the barrier and the interfaces.
A ``good'' N barrier is defined to have  sizeable $T_n$, if multiple pair
tunneling terms are to be observed. This requires a thin  barrier with slowly
decaying spin and  charge correlations\cite{prox}.
It is important to
note that multiple pair tunneling, i.e. the differential resistance peaks at  $
n>1 $, depends on strong   anharmonic  interactions between the hole pairs
and  magnons. {\em These  interactions are an  essential part of the pSO(5)
theory} as   modelled by $\cH^{int}$ in Eq.(\ref{pSO5}).

The junction's conductance is calculated
in the standard fashion\cite{mahan}: the bias voltage $V$
transforms  the left bosons $b_{h_L }
\to e^{i2eVt} b_{h_L }$, which yields time dependent
operators $\cA_n(t)$. The  current is calculated  by second order
perturbation theory in $\cH^{tun-mag}$ yielding
\bea
I &=&\sum_n 2ne   X_n^{ret} (2eV)    \nonumber\\
X_n^{ret} (\omega) &=& i\int_{0}^\infty dt
e^{i\omega t} \langle \left[ A_n^\nd (t) ,A_n^\dagger\right]\rangle
\eea
For  singular contributions $I^{sing}$, we  ignore superconducting
condensate fluctuations $b^\dagger_h-\langle b^\dagger_h\rangle$, which have a
smooth spectrum. Similarly, we ignore the frequency dependence of
$T_n(\omega)$.  Setting
$b_R^\dagger \to \langle b^\dagger_h\rangle  $ and $b_L^\dagger \to e^{i2eVt}
\langle b^\dagger_h\rangle $  leads to 
\bea
 I^{sing} &=&\sum_n 2ne \sum_{|q_x|\le \pi/d,|q_y|\le \pi/W}     \langle b^\dagger_h\rangle^{4n}
|T_n[ \bq] |^2 \nonumber\\
&& \times \Im \sum_{\omega} S(\bq, i\omega+2neV+i0^+)  S(-\bq,
i\omega)
\eea
where the barrier dimensions are $d\times W$ (see Fig.\ref{fig:diag}), and
$\sum_\omega$ is a Matsubara sum.

\begin{figure}[htbp]
\vspace*{13pt}
   \leavevmode \epsfxsize0.75\columnwidth
     \epsfbox{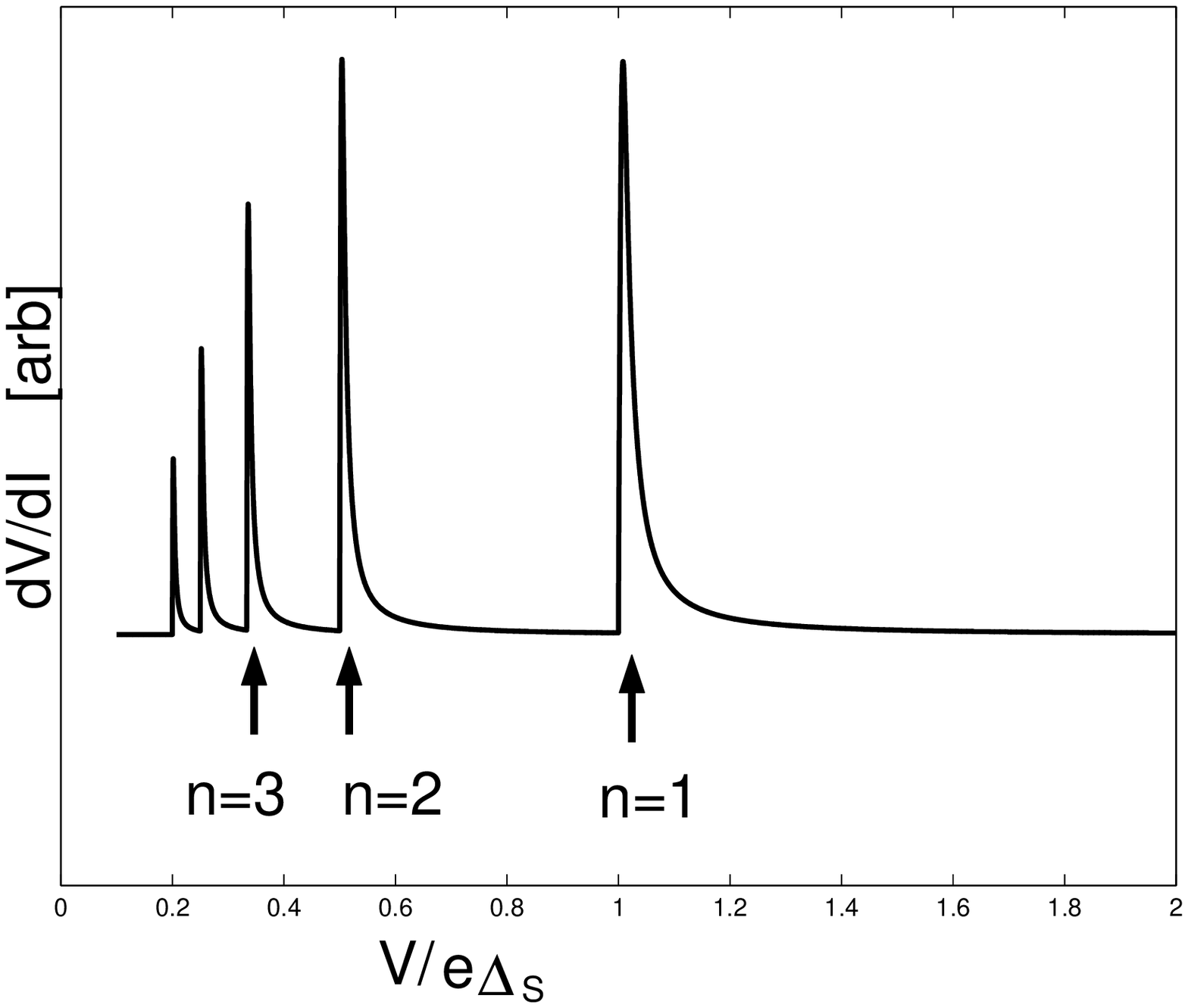}\vskip0.2pc
\fcaption{{\bf pSO(5)  Andreev peaks.}  Eq.
\ref{peaks} is plotted for a choice of $t_n/\Delta_s^{3/2}  = 2^{-n} 10^{-4}$,
$n\le 5$, and a  background conductance of unity. Below the $n$-th peak, the
excess tunneling charge is $2ne$, rather than BTK's $ne$.} \label{fig:peaks}
\end{figure}

For a nearly antiferromagnetic ``N'' barrier,
$T_n(\bx-\bx')$ in (\ref{H-tun}) decays  slowly  with the distance
between magnons.
Thus for a narrow barrier $d<<W$,  the magnons are excited at
$q_y\approx 0$, and the momentum sum reduces to a one dimensional sum over
$q_x$. At zero temperature we obtain
\bea
I^{sing} &=&\sum_n 2ne     \langle b^\dagger_h\rangle^{4n}
|T_n[ 0] |^2 \nonumber\\  &&
\times  s_0^2 \int {dq_x\over 2\pi}  {\delta(2neV-2\sqrt{c^2 q_x^2
+\Delta_s^2}) \over 2 (\Delta_s^2 + c^2 q_x^2)  }\nonumber\\
&\approx & \sum_n t_n {\theta(neV-\Delta_s)  \over  \Delta_s^{3/2}
\sqrt{ neV-\Delta_s } }
\label{peaks}
\eea
The last expression emphasizes the singular form of
$I^{sing}(V,\Delta_s)$ at the peaks.
For a large background  conductance $dI/dV>> dI^{sing} /dV$, the
inverse square root singularities in $I^{sing}$ create
peaks in the differential resistance $dV/dI$ at voltages
\be
V_n=\Delta_s/(ne),~~~~    n=1,2,\ldots ,~~~ Q_n= 2ne
\ee
where $Q_n$ is the excess tunneling charge below  the $n$-th
peak.  Note that $Q_n$ changes in increments of $2e$.
The differential resistance peak series is
depicted in Fig. \ref{fig:peaks}, for weak broadening of the singularities and
an arbitrary set of coefficients $t_n$.
 
\section{ Discussion, and Proposed Experiment}
\noindent
We have seen that magnon pair creation induces peaks in
the differential resistance which are similar {\em in appearance}  to the
Andreev peaks of the KBT mechanism.  The crucial difference is that  here
the singular  dissipative process  does {\em not} involve Cooper pair
breaking, but low energy antiferromagnetic  excitations. In the KBT mechanism,  a single sharp gap-like feature
can be obtained in a $d$ wave superconductor only by precise alignment of the $a$-$b$ axes of the two superconductors.

Here, one only requires the junction to be flat in the transverse direction,  such that $q_y$ is conserved
and the charge pairs
are coupled mostly to the one dimensional singularity of the magnon density of states. This requirement is less
stringent for weakly dispersive magnons near the resonance.

In KBT theory for two identical
superconductors, the peaks appear at voltages
 $V^{KBT}_n = 2\Delta/(ne),~~~~    n=1,2,\ldots$ which are the upper threshold
for tunneling of charges $Q_n=ne$. Thus,  KBT allows both even and odd number
of electron charges  to participate in the multiple Andreev reflection
process, as depicted in Fig. \ref{fig:btk}, while the pSO(5) theory expects only pair charges $Q_n=2ne$.

Observation of Andreev reflection enhanced shot noise $S(V)$ 
has been reported by Dieleman {\it et. al.}\cite{shot}
in a conventional SNS junction. 
They have measured  the tunneling charge  via the relation\cite{shot-theory}
$S=2Q_n I(V_n)$. The increment  of charge at the first Andreev peak at $2\Delta$ was clearly
seen to be of magnitude $e$.

We propose that a similar measurement in YBCO junctions could provide
a  decisive discrimination between the
processes of Fig.\ref{fig:btk} and Fig.\ref{fig:diag}.
The goal is to  measure the  charge {\em increments}
$Q_n-Q_{n-1}$  at any peak position $V_n$, $n=1,2,\ldots$
and see whether they are of magnitude $2e$ rather than $e$. 
The measurement would probably involve
a careful subtraction of the large but smooth
background quasiparticle contribution to the current and the noise
spectrum.  We eagerly look forward to results of such experiments.

\nonumsection{Acknowledgements}
\noindent
Support from the Israel Science Foundation
and the Fund for Promotion of Research at Technion is acknowledged.
We are grateful for the hospitality of the 
Institute for Theoretical Physics at
Santa Barbara, where this research was supported in part by the 
National Science Foundation under Grant
No. PHY94-07194.

\nonumsection{References}


\noindent
\begin{thebibliography}{000}
\bibitem{nmr}  H. Alloul, T. Ohno and P.Y. Mendels,    Phys. Rev  Lett. {\bf 63}. (1989).

\bibitem{tunn-exp}  L. Ozyuzer, J. F. Zasadzinski, N.
Miyakawa,   Int. J. Mod. Phys. {\bf B 29-31}, 3721,  (1999).
 
 \bibitem{arpes}  H. Ding {\it et. al}, Nature (London) {\bf 382}, 51054
(1996);
A. G. Loeser {\it et. al.},  Science {\bf 273}, 325 (1996).

 \bibitem{Deutscher} G. Deutscher, Nature (London) {\bf
397}, 40 (1999).

  \bibitem{phase-fluc}  V.J. Emery and S.A. Kivelson, 
Nature {\bf 374} 434-437 (1995)

\bibitem{pSO5} S.-C. Zhang,
J.-P. Hu, E. Arrigoni, W. Hanke and A. Auerbach,  Phys. Rev. {\bf B60}, 13070
(1999).

\bibitem{AA}  A. Auerbach and Ehud Altman,  cond-mat/0005420,
and Phys. Rev. Lett. (in press). 

\bibitem{ehud}  M. Havilio and A. Auerbach,  Phys. Rev. Lett. {\bf 83},
4848-4851 (1999).

\bibitem{comm-SO5} The projection breaks {\em quantum} SO(5) symmetry and
eliminates the high energy charge modes. SO(5) symmetry can only be retained
in the interaction $\cH^{int}$, and classical ground state manifold. The vacuum is
a projected RVB state and not an empty Fermi surface. These are some of the
crucial differences between the pSO(5) theory and the earlier  unprojected
SO(5) theory of S.-C. Zhang,  Science {\bf 275}, 1089-1096  (1997).
\bibitem{rvb}   P.W. Anderson, Science {\bf 235}, 1196 (1987).

\bibitem{moshe}  E. Altman and A. Auerbach,  unpublished.

\bibitem{uemura}   Y.J. Uemura,  Physica C {\bf 282-287}, 194-197 (1997).
 {\bf 235}, 1196 (1987).

\bibitem{Fong} H.F. Fong{\it et. al.},  Phys. Rev. Lett. {\bf 78}, 713 (1997);
cond-mat/9910041.


 \bibitem{kbt}   T.M. Klapwijk, G.E.  Blonder and  M. Tinkham,  Physica B,
{\bf 109\& 110}, 1657-1664 (1982); M. Octavio,    M. Tinkham,   G.E.  Blonder 
and  T.M. Klapwijk,  Phys. Rev. B {\bf 27}, 6739 (1983).

 \bibitem{nesher} O. Nesher and G. Koren,  Phys. Rev. B{\bf 60}, 1 (1999).


\bibitem{Racah} Racah, D. and G. Deutscher,  Physica C{\bf 2637},  218-224 
(1996).

\bibitem{prox}  R.S. Decca, H.D. Drew, E.Osquiguil, B. Mairov and J. Guimpel,
cond-mat/0003213

\bibitem{mahan} G. D. Mahan, {\em Many-particle physics}, Plenum (1986), Ch.
9.3.


\bibitem{shot}  P. Dieleman {\it et. al.},  Phys. Rev. Lett. {\bf 79}, 3486
(1997). 

\bibitem{shot-theory}  J.C. Cuevas, A. Martin-Rodero and A. Levy
Yeyati, 
Phys. Rev. Lett. {\bf 82}, 4086 (1999), and references therein.





\end{thebibliography}
\end{document}